\documentclass[epj]{svjour}

\usepackage[dvips]{graphicx}
\usepackage{array}
\usepackage{amssymb}
\usepackage{amsmath}
\usepackage{hhline}
\usepackage{longtable}
\usepackage{dcolumn}
\usepackage{bm}
\usepackage{subfigure}
\usepackage{epsfig}
\usepackage{latexsym,amsmath}
\usepackage{amsbsy}

\begin{document}
\title{Recursive weighted treelike networks}

\author{Zhongzhi Zhang\inst{1,2} \and Shuigeng Zhou\inst{1,2} \and Lichao Chen\inst{1,2} \and Jihong Guan\inst{3} \and Lujun Fang\inst{1,2} \and Yichao Zhang\inst{4}}                     
\offprints{zhangzz@fudan.edu.cn (Z. Z. Zhang)}          
\institute{Department of Computer Science and Engineering, Fudan
University, Shanghai 200433, China \and Shanghai Key Lab of
Intelligent Information Processing, Fudan University, Shanghai
200433, China \and Department of Computer Science and Technology,
Tongji University, 4800 Cao'an Road, Shanghai 201804, China \and
Material and Engineering Institute, Shanghai University, Shanghai
200072, China}

\date{Received: date / Revised version: date}

\abstract{We propose a geometric growth model for weighted
scale-free networks, which is controlled by two tunable parameters.
We derive exactly the main characteristics of the networks, which
are partially determined by the parameters. Analytical results
indicate that the resulting networks have power-law distributions of
degree, strength, weight and betweenness, a scale-free behavior for
degree correlations, logarithmic small average path length and
diameter with network size. The obtained properties are in agreement
with empirical data observed in many real-life networks, which shows
that the presented model may provide valuable insight into the real
systems.
} 

 \maketitle

\section{Introduction}
Complex networks~\cite{AlBa02,DoMe02,Ne03,BoLaMoChHw06,BoSaVe07}
describe a number of real-life systems in nature and society, such
as Internet~\cite{FaFaFa99}, World Wide Web~\cite{AlJeBa99},
metabolic networks~\cite{JeToAlOlBa00}, protein networks in the
cell~\cite{JeMaBaOl01}, worldwide airport
networks~\cite{BaBaPaVe04,LiCa04}, co-author networks
\cite{Ne01a,Newman01,BaJeNeRaScVi02,LiWuWaZhDiFa07} and sexual
networks \cite{LiEdAmStAb01}. Since the publication of the
pioneering papers by Watts and Strogatz on small-world networks
\cite{WaSt98} and Barab\'asi and Albert on scale-free networks
\cite{BaAl99}, modeling real-life systems has attracted an
exceptional amount of attention within the physics
community~\cite{AlBa02,DoMe02,Ne03,BoLaMoChHw06,BoSaVe07}.

Up to now, the research on modeling real-life systems has been
primarily focused on binary networks, i.e., edges among nodes are
either present or absent, represented as binary states. The purely
topological structure of binary networks, however, misses some
important attributes of real-world networks. Actually, many real
networked systems exhibit a large heterogeneity in the capacity and
the intensity of the connections, which is far beyond Boolean
representation. Examples include strong and weak ties between
individuals in social
networks~\cite{Ne01a,Newman01,BaJeNeRaScVi02,LiWuWaZhDiFa07}, the
varying interactions of the predator-prey in food networks~\cite
{KrFrMaUlTa03}, unequal traffic on the Internet~\cite{FaFaFa99} or
of the passengers in airline networks~\cite{BaBaPaVe04,LiCa04}.
These systems can be better described in terms of weighted networks,
where the weight on the edge provides a natural way to take into
account the connection strength. In the last few years, modeling
real systems as weighted complex networks has attracted an
exceptional amount of attention.

The first evolving weighted network model was proposed by Yook
\emph{et al.} (YJBT model)~\cite{YoJeBaTu01}, where the topology and
weight are driven by only the network connection based on
preferential attachment (PA) rule. In Ref.~\cite{ZhTrZhHu03}, a
generalized version of the YJBT model was presented, which
incorporates a random scheme for weight assignments according to
both the degree and the fitness of a node. In the YJBT model and its
generalization, edge weights are randomly assigned when the edges
are created, and remain fixed thereafter. These two models overlook
the possible dynamical evolution of weights occurring when new nodes
and edges enter the systems. On the other hand, the evolution and
reinforcements of interactions is a common characteristic of
real-life networks, such as airline
networks~\cite{BaBaPaVe04,LiCa04} and scientific collaboration
networks~\cite{Ne01a,Newman01,BaJeNeRaScVi02,LiWuWaZhDiFa07}. To
better mimic the reality, Barrat, Barth\'elemy, and Vespignani
introduced a model (BBV) for the growth of weighted networks that
couples the establishment of new edges and nodes and the weights'
dynamical evolution~\cite{BaBaVe04a,BaBaVe04b}. The BBV model is
based on a weight-driven dynamics~\cite{AnKr05} and on a weights'
reinforcement mechanism, it is the first weighted network model that
yields a scale-free behavior for the weight, strength, and degree
distributions. Enlightened by BBV's remarkable work, various
weighted network models have been proposed to simulate or explain
the properties found in real
systems~\cite{BoLaMoChHw06,BoSaVe07,WaWaHuYaQu05,WuXuWa05,GoKaKi05,MuMa06,XiWaWa07,LiWaFaDiWu06}.

While a lot of models for weighted networks have been presented,
most of them are stochastic~\cite{BoLaMoChHw06}. Stochasticity
present in previous models, while according with the major
properties of real-life systems, makes it difficult to gain a
visual understanding of how do different nodes relate to each
other forming complex weighted networks~\cite{BaRaVi01}. It would
therefore of major theoretical interest to build deterministic
weighted network models. Deterministic network models allow one to
compute analytically their features, which play a significant
role, both in terms of explicit results and as a guide to and a
test of simulated and approximate
methods~\cite{BaRaVi01,DoGoMe02,CoFeRa04,ZhRoZh07,JuKiKa02,RaSoMoOlBa02,RaBa03,AnHeAnSi05,DoMa05,ZhCoFeRo05,ZhRo05,Bobe05,BeOs79,HiBe06,CoOzPe00,CoSa02,ZhRoGo05,ZhRoCo05a}.
So far, the first and the only deterministic weighted network
model has been proposed by Dorogovtsev and Mendes
(DM)~\cite{DoMe05}. In the DM model, only the distributions of the
edge weight, of node degree and of the node strength are computed,
while other characteristics are omitted.

In this paper, we introduce a deterministic model for weighted
networks using a recursive construction. The model is controlled
by two parameters. We present an exhaustive analysis of many
properties of our model, and obtain the analytic solutions for
most of the features, including degree distributions, strength
distribution, weight distribution, betweenness distribution,
degree correlations, average path length, and diameter. The
obtained statistical characteristics are equivalent with some
random models (including BBV model).

\section{The model}

\begin{figure}[t]
  \centering\includegraphics[width=8cm]{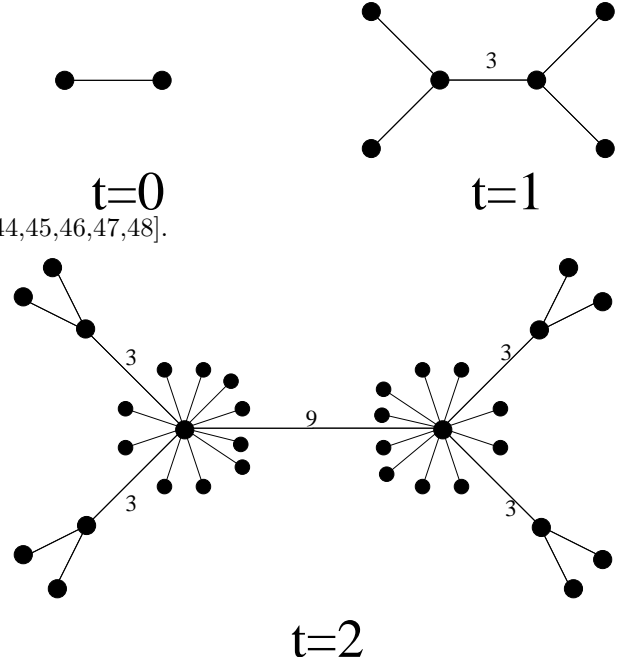}
  \caption{Illustration of a deterministically growing network in the
case of $m=2$  and  $\delta=1$, showing the first three steps of
growing process. The bare edges denote the edges of weight
1.}\label{recursive}
\end{figure}

The network, controlled by two parameters $m$ and $\delta$, is
constructed in a recursive way. We denote the network after $t$
steps by $Q(t)$, $t\geq 0$ (see Fig.~\ref{recursive}). Then the
network at step $t$ is constructed as follows. For $t=0$, $Q(0)$ is
an edge with unit weight connecting two nodes. For $t\geq 1$, $Q(t)$
is obtained from $Q(t-1)$. We add $mw$ ($m$ is positive integer) new
nodes for either end of each edge with weight $w$, and connect each
of $mw$ new nodes to one end of the edge by new edges of unit
weight; moreover, we increase weight $w$ of the edge by $m\delta w$
($\delta$ is positive integer). In the special case $\delta=0$, it
becomes binary networks, where all edges are
identical~\cite{JuKiKa02,GhOhGoKaKi04,Bobe05}.

 Let us consider
the total number of nodes $N_t$, the total number of edges $E_t$
and the total weight of all edges $W_t$ in $Q(t)$. Denote $n_v(t)$
as the number of nodes created at step $t$. Note that the addition
of each new node leads to only one new edge, so the number of
edges generated at step $t$ is $n_e(t)=n_v(t)$. By construction,
for $t\geq 1$, we have
\begin{equation}\label{nv01}
n_v(t)=2mW_{t-1},
\end{equation}
\begin{equation}\label{Et01}
E_t=E_{t-1}+n_v(t),
\end{equation}
and
\begin{equation}\label{Wt01}
W_t=W_{t-1}(1+m\delta)+2mW_{t-1}.
\end{equation}
On the right-hand side of Eq.~(\ref{Wt01}), the first item is the
sum of weigh of the old edges, and the second term describe the
total weigh of the new edges generated in step $t$. We can
simplify Eq.~(\ref{Wt01}) to yield
\begin{equation}\label{Wt02}
W_t=(1+m\delta+2m)W_{t-1}.
\end{equation}
Considering the initial condition $W_0=1$, we obtain
\begin{equation}\label{Wt03}
W_t=(1+m\delta+2m)^{t}.
\end{equation}
Substituting Eq.~(\ref{Wt03}) into Eq.~(\ref{nv01}), the number of
nodes created at step $t$ ($t\geq 1$) is obtained to be
\begin{equation}\label{nv02}
n_v(t)=2m(1+m\delta+2m)^{t-1}.
\end{equation}
Then the total number of nodes present at step $t$ is
\begin{align}\label{Nt01}
N_t &=\sum_{t_i=0}^{t}n_v(t_i)\nonumber\\
&=\frac{2}{2+\delta}\,\left[(1+m\delta+2m)^{t}+\delta +1\right].
\end{align}
Combining Eq.~(\ref{nv02}) with Eq.~(\ref{Et01}) and considering
$E_0=1$, it follows that
\begin{equation}\label{Et02}
E_t=\frac{2}{2+\delta}\,\left[(1+m\delta+2m)^{t}+\delta \right].
\end{equation}
Thus for large $t$, the average node degree $\overline{k}_t=
\frac{2E_t}{N_t}$ and average edge weight $\overline{w}_t=
\frac{W_t}{E_t}$ are approximately equal to $2$ and
$\frac{2+\delta}{2}$, respectively.

\section{Network properties}
Below we will find that the tunable parameters $m$ and $\delta$
control some relevant characteristics of the weighted network
$Q(t)$. We focus on the weight distribution, strength
distribution, degree distribution, degree correlations,
betweenness distribution, average path length, and diameter.

\subsection{Weight distribution}
Note that all the edges emerging simultaneously have the same
weight. Let $w_e(t)$ be the weight of edge $e$ at step $t$. Then
by construction, we can easily have
\begin{equation}\label{we01}
w_e(t)=(1+m\delta)w_e(t-1).
\end{equation}
If edge $e$ enters the network at step $\tau$, then $w_e(\tau)=1$.
Thus
\begin{equation}\label{we02}
w_e(t)=(1+m\delta)^{t-\tau}.
\end{equation}
Therefore, the weight spectrum of the network is discrete. It
follows that the weight distribution is given by
\begin{equation}
P(w)=\left\{\begin{array}{lc} {\displaystyle{n_e(0)\over
E_t}={\delta+2\over 2[(1+m\delta+2m)^{t}+\delta ]} }
& \ \hbox{for}\ \tau=0,\\
{\displaystyle{n_e(\tau)\over
E_t}={m(2+\delta)\,(1+m\delta+2m)^{\tau-1}\over
 (1+m\delta+2m)^{t}+\delta} }
& \ \hbox{for}\  \tau\ge 1,\\
0 & \ \hbox{otherwise}\end{array} \right.
\end{equation}
and that the cumulative weight distribution \cite{Ne03,DoGoMe02}
is
\begin{align}
P_{\rm cum}(w)&=\sum_{\mu \leq \tau}\frac{n_e(\mu)}{E_t}\nonumber\\ &={(1+m\delta+2m)^{\tau}+\delta \over
(1+m\delta+2m)^{t}+\delta }.
\end{align}
Substituting for $\tau$ in this expression using $\tau=t-\frac{\ln
w}{\ln (1+m\delta)}$ gives
\begin{align}\label{gammaw}
P_{\rm cum}(w)&={(1+m\delta+2m)^{t} w^{-\frac{\ln(1+m\delta+2m)}{\ln(1+m\delta)}}+\delta \over (1+m\delta+2m)^{t}+\delta}\nonumber\\
          &\approx w^{-\frac{\ln(1+m\delta+2m)}{\ln(1+m\delta)}}\qquad \qquad \hbox{for large $t$}.
\end{align}
So the weight distribution follows a power law with the exponent
$\gamma_{w}=1+{\frac{\ln(1+m\delta+2m)}{\ln(1+m\delta)}}$.

\subsection{Strength distribution}
In a weighted network, a node strength is a natural genearlization
of its degree. The strength $s_{i}$ of node $i$ is defined as
\begin{equation}\label{si01}
s_i = \sum_{j\in \Omega_{i}} w_{ij} \, ,
\end{equation}
where $w_{ij}$ denotes the weight of the edge between nodes $i$
and $j$, $\Omega_{i}$ is the set of all the nearest neighbors of
$i$. The strength distribution $P(s)$ measures the probability
that a randomly selected node has exactly strength $s$.

Let $s_i(t)$ be the strength of node $i$ at step $t$. If node $i$
is added to the network at step $t_i$, then $s_i(t_i)=1$.
Moreover, we introduce the quantity $\Delta s_i(t)$, which is
defined as the difference between $s_i(t)$ and $s_i(t-1)$. By
construction, we can easily obtain
\begin{align}\label{Deltasi01}
\Delta s_i(t)&=s_i(t)-s_i(t-1)\nonumber\\
&=m\delta\sum_{j\in \Omega_{i}}w_{ij}+m\sum_{j\in \Omega_{i}} w_{ij}\nonumber\\
&=m\delta s_i(t-1)+ m\,s_i(t-1).
\end{align}
Here the first item accounts for the increase of weight of the old
edges incident with $i$, which exist at step $t-1$. The second
term describe the total weigh of the new edges with unit weight
that are generated at step $t$ and connected to node $i$.

From Eq.~(\ref{Deltasi01}), we can derive following recursive
relation:
\begin{equation}\label{si02}
s_i(t)=(1+m\delta+m)s_i(t-1).
\end{equation}
Using $s_i(t_i)=1$, we obtain
\begin{equation}\label{si03}
s_i(t)=(1+m\delta+m)^{t-t_i}.
\end{equation}
Since the strength of each node has been obtained explicitly as in
Eq.~(\ref{si03}), we can get the strength distribution via its
cumulative distribution~\cite{Ne03,DoGoMe02}, i.e.
\begin{align}\label{pcums01}
P_{\rm cum}(s)&=\sum_{\mu \leq t_i}\frac{n_v(\mu)}{N_t}\nonumber\\
&={(1+m\delta+2m)^{t_i}+\delta +1 \over (1+m\delta+2m)^{t}+\delta
+1}.
\end{align}
From Eq.~(\ref{si03}), we can derive $t_i=t-\frac{\ln s}{\ln
(1+m\delta+m)}$. Substituting the obtained result of $t_i$ into
Eq.~(\ref{pcums01}) gives
\begin{eqnarray}\label{gammas}
P_{\rm cum}(s)&=&{(1+m\delta+2m)^{t}\, s^{-\frac{\ln(1+m\delta+2m)}{\ln(1+m\delta+m)}}+\delta +1 \over (1+m\delta+2m)^{t}+\delta+1}\nonumber\\
          &\approx& s^{-\frac{\ln(1+m\delta+2m)}{\ln(1+m\delta+m)}} \qquad \qquad \hbox{for large $t$}.
\end{eqnarray}
Thus, node strength distribution exhibits a power law behavior
with the exponent
$\gamma_{s}=1+{\frac{\ln(1+m\delta+2m)}{\ln(1+m\delta+m)}}$.

\subsection{Degree distribution}
The most important property of a node is the degree, which is
defined as the number of edges incident with the node. Similar to
strength, in our model, all simultaneously emerging nodes have the
same degree.  Let $k_i(t)$ be the degree of node $i$ at step $t$.
If node $i$ is added to the graph at step $t_i$, then by
construction $k_i(t_i)=1$. After that, the degree $k_i(t)$ evolves
as
\begin{equation}\label{ki01}
k_i(t)=k_i(t-1)+m\,s_{i}(t-1),
\end{equation}
where $ms_{i}(t-1)$ is the degree increment $\Delta k_i(t)$ of
node $i$ at step $t$. Substituting Eq.~(\ref{si03}) into
Eq.~(\ref{ki01}), we have
\begin{equation}\label{deltaki01}
\Delta k_i(t)=m\,(1+m\delta+m)^{t-1-t_i}.
\end{equation}
Then the degree $ k_i(t)$ of node $i$ at time $t$ is
\begin{align}\label{ki02}
k_i(t)&=k_i(t_i)+\sum_{\eta=t_i+1}^{t}{\Delta
k_i(\eta)}\nonumber\\
&=\frac{(m\delta+1+m)^{t-t_i}+\delta}{\delta+1}.
\end{align}
Analogously to computation of cumulative strength distribution, one
can find the cumulative degree distribution
\begin{align}\label{gammak}
P_{\rm cum}(k)
&={(1+m\delta+2m)^{t}\, [(\delta+1)\,k-\delta]^{-\frac{\ln(1+m\delta+2m)}{\ln(1+m\delta+m)}}\over (1+m\delta+2m)^{t}+\delta+1}\nonumber\\
&\quad + {\delta +1 \over (1+m\delta+2m)^{t}+\delta+1}\nonumber\\
&\approx
[(\delta+1)\,k]^{-\frac{\ln(1+m\delta+2m)}{\ln(1+m\delta+m)}}
\qquad  \hbox{for large $t$}.
\end{align}
Thus, the degree distribution is scale-free with the same exponent
as $\gamma_{s}$, that is
$\gamma_{k}=\gamma_{s}=1+{\frac{\ln(1+m\delta+2m)}{\ln(1+m\delta+m)}}$.

\subsection{Betweenness distribution}
Betweenness of a node is the accumulated fraction of the total
number of shortest paths going through the given node over all
node pairs~\cite{Fr77,Newman01}. More precisely, the betweenness
of a node $i$ is
\begin{equation}
b_{i}=\sum_{j \ne i \neq k}\frac{\sigma_{jk}(i)}{\sigma_{jk}},
\end{equation}
where $\sigma_{jk}$ is the total number of shortest path between
node $j$ and $k$, and $\sigma_{jk}(i)$ is the number of shortest
path running through node $i$.

Since the considered network here is a tree, for each pair of
nodes there is a unique shortest path between
them~\cite{SzMiKe02,BoRi04,GhOhGoKaKi04}. Thus the betweenness of
a node is simply given by the number of distinct shortest paths
passing through the node. From Eqs.~(\ref{deltaki01}) and
(\ref{ki02}), we can easily derive that for $\alpha < \theta$ the
number of nodes with degree
$\frac{(m\delta+1+m)^{\alpha}+\delta}{\delta+1}$ which are direct
children of a node with degree
$\frac{(m\delta+1+m)^{\theta}+\delta}{\delta+1}$ is
$m(1+m\delta+m)^{\tau-1-\alpha}$. Then at time $t$, the
betweenness of a $\theta$-generation-old node $v$, which is
created at step $t-\theta$, denoted as $b_{t}(\theta)$ becomes
\begin{align}\label{between01}
b_{t}(\theta) &= \mathcal{C}_{t}^{\theta}\,\left[N_{t} -
\left(\mathcal{C}_{t}^{\theta}+1\right)\right]+\binom
{\mathcal{C}_{t}^{\theta}}{2}\nonumber \\
  &\quad- \sum_{\alpha=1}^{\theta-1}m(1+m\delta+m)^{\tau-1-\alpha} \binom
{\mathcal{C}_{t}^{\alpha}+1}{2},
\end{align}
where $\mathcal{C}_{t}^{\theta}$ denotes the total number of
descendants of node $v$ at time $t$, where the descendants of a node
are its children, its children's children, and so on. Note that the
descendants of node $v$ exclude $v$ itself. The first term in
Eq.~(\ref{between01}) counts shortest paths from descendants of $v$
to other vertices. The second term accounts for the shortest paths
between descendants of $v$. The third term describes the shortest
paths between descendants of $v$ that do not pass through $v$.

To find $b_{t}(\theta)$, it is necessary to explicitly determine
the descendants $\mathcal{C}_{t}^{\theta}$ of node $v$, which is
related to that of $v's$ children via~\cite{GhOhGoKaKi04}
\begin{equation}\label{child01}
\mathcal{C}_{t}^{\theta}=
\sum_{\alpha=1}^{\theta}m(1+m\delta+m)^{\alpha-1}
\left(\mathcal{C}_{t}^{\tau-\alpha}+1\right).
\end{equation}
Using $\mathcal{C}_{t}^{0}=0$, we can solve Eq.~(\ref{child01})
inductively,
\begin{equation}\label{child02}
\mathcal{C}_{t}^{\theta}= \frac{1}{\delta+2}\left [
(m\delta+1+2m)^{\tau}-1\right ].
\end{equation}
Substituting the result of Eq.~(\ref{child02}) and~(\ref{Nt01})
into Eq.~(\ref{between01}), we have
\begin{equation}\label{between02}
b_{t}(\theta) \simeq \frac{2}{(\delta+2)^{2}}\,
(m\delta+1+2m)^{t+\tau}.
\end{equation}
Then the cumulative betweenness distribution is
\begin{align}\label{pcumb01}
P_{\rm cum}(b)&=\sum_{\mu \leq
t-\tau}\frac{n_v(\mu)}{N_t}\nonumber\\
&={(1+m\delta+2m)^{t-\tau}+\delta +1 \over
(1+m\delta+2m)^{t}+\delta
+1}\nonumber \\
&\simeq {(1+m\delta+2m)^{t} \over (1+m\delta+2m)^{t+\tau}}\sim
{N_{t} \over b}\sim b^{-1},
\end{align}
which shows that the betweenness distribution exhibits a power law
behavior with exponent $\gamma_{b}=2$, the same scaling has been
also obtained for the $m=1$ case of the Barab\'asi-Albert (BA)
model describing a random scale-free treelike
network~\cite{SzMiKe02,BoRi04}.

\subsection{Degree correlations}
Degree correlation is a particularly interesting subject in the
field of network
science~\cite{MsSn02,PaVaVe01,VapaVe02,Newman02,Newman03c,ZhZh07},
because it can give rise to some interesting network structure
effects. An interesting quantity related to degree correlations is
the average degree of the nearest neighbors for nodes with degree
$k$, denoted as $k_{\rm nn}(k)$ \cite{PaVaVe01,VapaVe02}. When
$k_{\rm nn}(k)$ increases with $k$, it means that nodes have a
tendency to connect to nodes with a similar or larger degree. In
this case the network is defined as assortative
\cite{Newman02,Newman03c}. In contrast, if $k_{\rm nn}(k)$ is
decreasing with $k$, which implies that nodes of large degree are
likely to have near neighbors with small degree, then the network
is said to be disassortative. If correlations are absent, $k_{\rm
nn}(k)=const$.

We can exactly calculate $k_{\rm nn}$ for the networks using
Eq.~(\ref{ki02}) to work out how many links are made at a
particular step to nodes with a particular degree. We place
emphasis on the particular case of $\delta=0$. Except for the
initial two nodes generated at step 0, no nodes born in the same
step, which have the same degree, will be linked to each other.
All links to nodes with larger degree are made at the creation
step, and then links to nodes with smaller degree are made at each
subsequent steps. This results in the
expression~\cite{ZhRoZh07,DoMa05} for $k=(m+1)^{t-t_i}$
\begin{align}\label{knn01}
k_{\rm nn}(k)&={1\over n_v(t_i) k(t_i,t)} \nonumber\\
&\quad \Bigg(
  \sum_{t'_i=0}^{t'_i=t_i-1} m\cdot n_v(t'_i) k(t'_i,t_i-1)k(t'_i,t)\nonumber\\
  &\quad+\sum_{t'_i=t_i+1}^{t'_i=t} m\cdot n_v(t_i) k(t_i,t'_i-1)
  k(t'_i,t)\Bigg),
\end{align}
where $k(t_i,t)$ represents the degree of a node at step $t$,
which was generated at step $t_i$. Here the first sum on the
right-hand side accounts for the links made to nodes with larger
degree (i.e.\ $t'_i<t_i$) when the node was generated at $t_i$.
The second sum describes the links made to the current smallest
degree nodes at each step $t'_i>t_i$.

Substituting Eqs.~(\ref{nv02}) and (\ref{ki02}) into
Eq.~(\ref{knn01}), after some algebraic manipulations,
Eq.~(\ref{knn01}) is simplified to
\begin{align} \label{knn02}
k_{\rm nn}(k)&= \frac{2m+1}{m}\,\left [\frac{(m+1)^{2}}{2m+1}
\right]^{t_i}\nonumber\\
&\quad-\frac{m+1}{m}+\frac{m}{m+1}\,(t-t_i).
\end{align}
Thus after the initial step $k_{\rm nn}$ grows linearly with time.

Writing Eq. (\ref{knn02}) in terms of $k$, it is straightforward
to obtain
\begin{align} \label{knn3}
k_{\rm nn}(k)&= \frac{2m+1}{m}\,\left [\frac{(m+1)^{2}}{2m+1}
\right ]^{t}\,k^{-\frac{\ln\left
[\frac{(m+1)^{2}}{2m+1}\right ]}{\ln(m+1)}}\nonumber\\
&\quad-\frac{m+1}{m}+\frac{m}{m+1}\,\frac{\ln k}{\ln(m+1)}.
\end{align}
Therefore, $k_{\rm nn}(k)$ is approximately a power law function
of $k$ with negative exponent, which shows that the networks are
disassortative. Note that $k_{\rm nn}(k)$ of the Internet exhibit
a similar power-law dependence on the degree $k_{\rm nn}(k)\sim
k^{-\omega}$, with $\omega=0.5$ \cite{PaVaVe01}. Additionally, one
can easily check that for other values of $\delta>0$, the networks
will again be disassortative with respect to degree because of the
lack of connections between nodes with the same degree.

\subsection{Average path length}
Most real-life systems are small-world, i.e., they have a
logarithmic average path length (APL) with the number of their
nodes. Here APL means the minimum number of edges connecting a
pair of nodes, averaged over all node pairs. For general $m$ and
$\delta$, it is not easy to derive a closed formula for the
average path length of $Q(t)$. However, for the particular case of
$m=1$ and $\delta=0$, the network has a self-similar structure,
which allows one to calculate the APL analytically.

For simplicity, we denote the limiting network ($m=1$ and
$\delta=0$) after $t$ generations by $Q_{t}$. Then the average path
length of $Q_t$ is defined to be:
\begin{equation}\label{eq:app4}
  \bar{d}_t  = \frac{D_t}{N_t(N_t-1)/2}\,.
\end{equation}
In Eq.~(\ref{eq:app4}), $D_t$ denotes the sum of the total
distances between two nodes over all pairs, that is
\begin{equation}\label{eq:app5}
  D_t = \sum_{i,j \in Q_t} d_{i,j}\,,
\end{equation}
where $d_{i,j}$ is the shortest distance between node $i$ and $j$.

\begin{figure}[t]
  \centering\includegraphics[width=9cm]{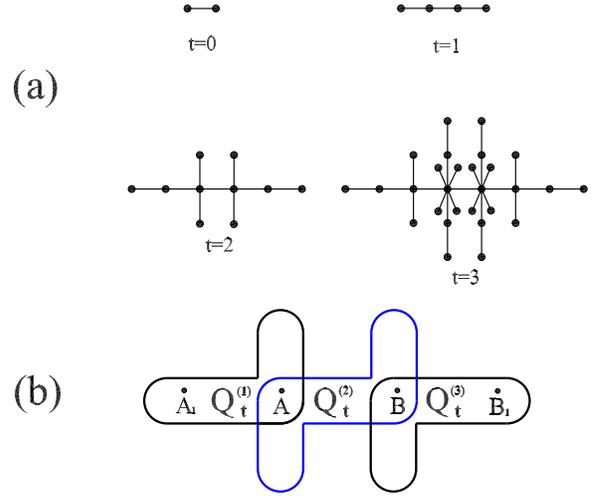}
  \caption{(Color online) (a) The first four steps of binary network growth for the limiting case of $m=1$ and
$\delta=0$ are shown. (b) The network after $t+1$
    generation, $Q_{t+1}$, can be obtained by joining three copies of generations $t$
    (i.e. $Q_t^{(1)}, Q_t^{(2)}, Q_t^{(3)}$) at the two hub nodes of highest degree, denoted by $A$ and $B$.}\label{apfig2}
\end{figure}

We can exactly calculate $\bar{d}_t$ according to the self-similar
network structure~\cite{HiBe06}. As shown in Fig.~\ref{apfig2}, the
network $Q_{t+1}$ may be obtained by joining at the hubs (the most
connected nodes) three copies of $Q_t$, which we label
$Q_t^{(\alpha)}$, $\alpha=1,2,3$~\cite{Bobe05,CoRobA05}. Then one
can write the sum over all shortest paths $D_{t+1}$ as
\begin{equation}\label{eq:app6}
  D_{t+1} = 3D_t + \Delta_t\,,
\end{equation}
where $\Delta_t$ is the sum over all shortest paths whose
endpoints are not in the same $Q_t$ branch. The solution of
Eq.~(\ref{eq:app6}) is
\begin{equation}\label{eq:app8}
  D_t = 3^{t-1} D_1 + \sum_{\tau=1}^{t-1} 3^{t-\tau-1} \Delta_\tau\,.
\end{equation}
The paths that contribute to $\Delta_t$ must all go through at
least either of the two hubs ($A$ and $B$) where the three
different $Q_t$ branches are joined. Below we will derive the
analytical expression for $\Delta_t$ named the crossing paths,
which is given by
\begin{equation}\label{APL31}
\Delta_t=\Delta_t^{1,2} + \Delta_t^{2,3} + \Delta_t^{1,3}\,,
\end{equation}
where $\Delta_t^{\alpha,\beta}$ denotes the sum of all shortest
paths with endpoints in $Q_t^{(\alpha)}$ and $Q_t^{(\beta)}$. If
$Q_t^{(\alpha)}$ and $Q_t^{(\beta)}$ meet at an edge node,
$\Delta_t^{\alpha,\beta}$ rules out the paths where either
endpoint is that shared edge node. If $Q_t^{(\alpha)}$ and
$Q_t^{(\beta)}$ do not meet, $\Delta_t^{\alpha,\beta}$ excludes
the paths where either endpoint is any edge node.

By symmetry, $\Delta_t^{1,2} = \Delta_t^{2,3}$, so that
\begin{equation}\label{APL32}
\Delta_t = 2 \Delta_t^{1,2}+\Delta_t^{1,3}\,,
\end{equation}
where $\Delta_t^{1,2}$ and $\Delta_t^{1,3}$ are given by the sum
\begin{align}\label{app12}
  \Delta_t^{1,2} &= \sum_{\substack{i \in Q_t^{(1)},\,\,j\in
      Q_t^{(2)}\\ i,j \ne A}} d_{i,j}
\end{align}
and
\begin{align}\label{app12}
  \Delta_t^{1,3} &= \sum_{\substack{i \in Q_t^{(1)},\,\,j\in
      Q_t^{(3)}\\ i \ne A,\,\,j\ne B}} d_{i,j},
\end{align}
respectively. In order to find $\Delta_t^{1,2}$ and
$\Delta_t^{1,3}$, we define
\begin{align}
d_t^\text{tot} &\equiv \sum_{Z \in Q_t^{(2)}}d_{Z,A}\, ,\nonumber\\
d_t^\text{near} &\equiv \sum_{\substack{Z \in Q_t^{(2)}\\ d_{Z,A}
<
    d_{Z,B}}} d_{Z,A}\,,\qquad N_t^\text{near} \equiv \sum_{\substack{Z
    \in Q_t^{(2)}\\ d_{Z,A} < d_{Z,B}}} 1\,,\nonumber\\
d_t^\text{far} &\equiv \sum_{\substack{Z \in Q_t^{(2)}\\ d_{Z,A} >
    d_{Z,B}}} d_{Z,A}\,,\qquad N_t^\text{far} \equiv \sum_{\substack{Z
    \in Q_t^{(2)}\\ d_{Z,A} > d_{Z,B}}} 1\,,
\label{eq:app32}
\end{align}
where $Z\neq A$. Since $A$ and $B$ are linked by one edge, for any
node $i$ in the network, $d_{i,A}$ and $d_{i,B}$ can differ by at
most 1, then we can easily have $d_t^\text{tot} = d_t^\text{near}
+ d_t^\text{far}$ and $N_t = N_t^\text{near} + N_t^\text{far} +
1$. By symmetry $N_t^\text{near}+1 = N_t^\text{far}$. Thus, by
construction, we obtain
\begin{equation}
N_t= 2\,(N_t^\text{near} +1).
\end{equation}
Combining this with Eq.~(\ref{Nt01}), we obtain partial quantities
in Eq.~\eqref{eq:app32} as
\begin{equation}\label{APL36}
  N_t^\text{far} -1= N^\text{near}_t = \frac{1}{2}
  \left(3^t-1\right).
\end{equation}
Now we return to the quantity $\Delta_t^{1,2}$ and
$\Delta_t^{1,3}$, both of which can be further decomposed into the
sum of four terms as
\begin{align}
\Delta_t^{1,2} &= \sum_{\substack{i \in Q_t^{(1)},\,\,j\in
      Q_t^{(2)}\\ i,j \ne A}} d_{i,j} \nonumber\\
&= \sum_{\substack{i \in Q_t^{(1)},\,\,j\in
      Q_t^{(2)},\,\,i,j \ne A \\ d_{i,A} > d_{i,A_{1}},\,\, d_{j,A} > d_{j,B}}}
  (d_{i,A}+d_{j,A})\nonumber\\
&\quad + \sum_{\substack{i \in Q_t^{(1)},\,\,j\in
      Q_t^{(2)},\,\,i,j \ne A \\ d_{i,A} < d_{i,A_{1}},\,\, d_{j,A} > d_{j,B}}}
  (d_{i,A}+d_{j,A})\nonumber\\
&\quad +\sum_{\substack{i \in Q_t^{(1)},\,\,j\in
      Q_t^{(2)},\,\,i,j \ne A \\ d_{i,A} > d_{i,A_{1}},\,\, d_{j,A} < d_{j,B}}}
  (d_{i,A}+d_{j,A})\nonumber\\
&\quad  + \sum_{\substack{i \in Q_t^{(1)},\,\,j\in
      Q_t^{(2)},\,\,i,j \ne A \\ d_{i,A} < d_{i,A_{1}},\,\, d_{j,A} < d_{j,B}}}
  (d_{i,A}+d_{j,A})\nonumber\\
&= 2(N_t-1)(d_t^\text{near}+d_t^\text{far})\,,\label{APL37}
\end{align}
and
\begin{align}
\Delta_t^{1,3} &= \sum_{\substack{i \in Q_t^{(1)},\,\,j\in
      Q_t^{(3)}\\ i \ne A,\,\,j \ne B}} d_{i,j} \nonumber\\
&= \sum_{\substack{i \in Q_t^{(1)},\,\,j\in
      Q_t^{(3)},\,\,i \ne A,\,\,j \ne B \\ d_{i,A} > d_{i,A_{1}},\,\, d_{j,B} > d_{j,B_{1}}}}
  (d_{i,A}+d_{j,A}+1)\nonumber\\
&\quad + \sum_{\substack{i \in Q_t^{(1)},\,\,j\in
      Q_t^{(3)},\,\,i \ne A,\,\,j \ne B \\ d_{i,A} < d_{i,A_{1}},\,\, d_{j,B} > d_{j,B_{1}}}}
  (d_{i,A}+d_{j,A}+1)\nonumber\\
&\quad +\sum_{\substack{i \in Q_t^{(1)},\,\,j\in
      Q_t^{(3)},\,\,i \ne A,\,\,j \ne B \\ d_{i,A} > d_{i,A_{1}},\,\, d_{j,B} < d_{j,B_{1}}}}
  (d_{i,A}+d_{j,A}+1)\nonumber\\
&\quad  + \sum_{\substack{i \in Q_t^{(1)},\,\,j\in
      Q_t^{(3)},\,\,i \ne A,\,\,j \ne B \\ d_{i,A} < d_{i,A_{1}},\,\, d_{j,B} < d_{j,B_{1}}}}
  (d_{i,A}+d_{j,A}+1)\nonumber\\
&= 2(N_t-1)(d_t^\text{near}+d_t^\text{far})+(N_t-1)^2\,,
\label{APL38}
\end{align}
respectively. Having $\Delta_n^{1,2}$ and $\Delta_n^{1,3}$ in
terms of the quantities in Eq.~\eqref{eq:app32}, the next step is
to explicitly determine these quantities unresolved.

Considering the self-similar structure of the graph, we can easily
know that at time $t+1$, the quantities $d_{t+1}^\text{near}$ and
$d_{t}^\text{far}$ are related to each other, both of which evolve
as
\begin{equation}
\left\{\begin{array}{lc} {\displaystyle{d^\text{near}_{t+1} =
d^\text{far}_t+2\,d^\text{near}_t\, ,} }\\
{\displaystyle{d^\text{far}_{t} = d^\text{near}_t+N_t^\text{far}\,.} }\\
\end{array} \right.
\end{equation}
From the two recursive equations we can obtain
\begin{equation}\label{APL39}
\left\{\begin{array}{lc}
{\displaystyle{d^\text{near}_{t} = \frac{1}{12} \left(-3+3^{1+t}+2t\cdot3^t\right)\, ,} }\\
{\displaystyle{d^\text{far}_{t} = \frac{1}{12} \left(3+3^{2+t}+2 t\cdot3^t\right)\,.} }\\
\end{array} \right.
\end{equation}
Substituting the obtained expressions in Eqs.~\eqref{APL36} and
\eqref{APL39} into Eqs.~\eqref{APL37},~\eqref{APL38}
and~\eqref{APL32}, the crossing paths $\Delta_t$ is obtained to be
\begin{equation}\label{APL10}
  \Delta_t = 7\cdot 9^{t}+2t\cdot9^{t}.
\end{equation}
Inserting Eq.~\eqref{APL10} into Eq.~\eqref{eq:app8} and using
$D_1 = 10$, we have
\begin{equation}\label{APL11}
  D_t = 3^{-1+t} \left(1+2\cdot3^t+ t\cdot3^t\right).
\end{equation}
Substituting Eqs.~(\ref{Nt01}) and (\ref{APL11}) into
(\ref{eq:app4}), the exact expression for the average path length
is obtained to be
\begin{equation}\label{eq:app10}
  \bar{d}_t = \frac{2 \left(1+2\cdot3^t+ t\cdot3^t\right)}{3 \left(1+3^t\right)}.
\end{equation}
In the infinite network size limit ($t \rightarrow \infty$),
\begin{equation}\label{eqapp56}
\bar{d}_{t} \simeq \frac{2}{3}\,t+\frac{4}{3} \sim \ln N_{t},
\end{equation}
which means that the average path length shows a logarithmic
scaling with the size of the network.

\subsection{Diameter}
Although we do not give a closed formula of APL of $Q(t)$ for
general $m$ and $\delta$ in the previous subsection, here we will
provide the exact result of the diameter of $Q(t)$ denoted by
$Diam(Q(t))$ for all $m$ and $\delta$, which is defined as the
maximum of the shortest distances between all pairs of nodes.
Small diameter is consistent with the concept of small-world. The
obtained diameter scales logarithmically with the network size.
Now we present the computation details as follows.

Clearly, at step $t=0$, $Diam(Q(0))$ is equal to 1. At each step
$t\geq 1$, we call newly-created nodes at this step \emph{active
nodes}. Since all active nodes are attached to those nodes existing
in $Q(t-1)$, so one can easily see that the maximum distance between
arbitrary active node and those nodes in $Q(t-1)$ is not more than
$Diam(Q(t-1))+1$ and that the maximum distance between any pair of
active nodes is at most $Diam(Q(t-1))+2$. Thus, at any step, the
diameter of the network increases by 2 at most. Then we get $2(t+1)$
as the diameter of $Q(t)$. Note that the logarithm of the size of
$Q(t)$ is approximately equal to $t\ln (1+m\delta+2m)$ in the limit
of large $t$. Thus the diameter is small, which grows
logarithmically with the network size.

\section{Conclusion}
In summary, we have introduced and investigated a deterministic
weighted network model in a recursive fashion, which couples
dynamical evolution of weight with topological network growth. In
the process of network growth, edges with large weight gain more new
links, which occurs in many real-life networks, such as scientific
collaboration
networks~\cite{DoMe05,Ne01a,Newman01,BaJeNeRaScVi02,LiWuWaZhDiFa07}.
We have obtained the exact results for the major properties of our
model, and shown that it can reproduce many features found in real
weighted networks as the famous BBV
model~\cite{BaBaVe04a,BaBaVe04b}. Our model can provide a visual and
intuitional scenario for the shaping of weighted networks. We
believe that our study could be useful in the understanding and
modeling of real-world networks.

\section*{Acknowledgment}
This research was supported by the National Natural Science
Foundation of China under Grant Nos. 60496327, 60573183, and
90612007, and the Postdoctoral Science Foundation of China under
Grant No. 20060400162. 


\end{document}